\documentclass{article}
\usepackage{latexsym}
\usepackage{graphicx}
\usepackage{epsfig}
\newcommand \be {\begin{equation}}
\newcommand \bea {\begin{eqnarray}}
\newcommand \ee {\end{equation}}
\newcommand \eea {\end{eqnarray}}

\newcommand{\bit}{\begin{itemize}}
\newcommand{\eit}{\end{itemize}}

\newcommand{\eps}{\epsilon}

\begin{document}
\topskip 2cm

\begin{titlepage}

\rightline{hep-ph/0411197}

\rightline{MIT-CTP-3566}

\bigskip

\begin{center}
{\Large\bf On the Vacuum Cherenkov Radiation in Noncommutative} \\

\bigskip
\bigskip

{\Large \bf Electrodynamics and the Elusive Effects of Lorentz Violation} \\

\vspace{2.5cm}

{\large Paolo Castorina$^{1}$, Alfredo Iorio$^{2,3}$, Dario Zappal\`a$^{1}$} \\
\vspace{.5cm}
{\sl $^{1}$ Department of Physics, University of Catania and INFN, Sezione di Catania} \\
{\sl Citt\`a Universitaria,  Via S. Sofia 64, I-95123 Catania, Italy}\\
{\sl $^{2}$ Center for Theoretical Physics, Massachusetts Institute of Technology}\\
{\sl 77, Massachusetts Avenue, Cambridge MA, U.S.A.} \\
{\sl $^{3}$ Department of Physics, University of Salerno and INFN, Gruppo collegato di Salerno} \\
{\sl Via Salvador Allende, Baronissi (SA), Italy} \\
\vspace{.5cm}
{E-mail: paolo.castorina@ct.infn.it;  iorio@lns.mit.edu;  dario.zappala@ct.infn.it}\\
\vspace{.5cm}

\vspace{2.5cm}

\begin{abstract}
\noindent We show that in the framework of noncommutative
classical electrodynamics Cherenkov radiation is permitted in
vacuum and we explicitly compute its spectrum at first order in
the noncommutative parameter. We discuss the phenomenological
impact of the merge of this new analysis with the old results of
the substantial modification to the spectrum of the synchrotron
radiation obtained in \cite{Castorina:2002vs}. We propose to
consider the pulsars' radiation spectrum - due to its very strong
magnetic field - to investigate these Lorentz violating effects in
astrophysical phenomena.
\end{abstract}
\end{center}

\vfill


\end{titlepage}

\newpage

\section{Introduction}

The possible occurrence of noncommutative spatiotemporal
coordinates has been extensively investigated in physics
\cite{Doplicher:1994tu} and mathematics \cite{connes}.
Noncommutative perturbative {\it quantum} field theories seem
unappealing for phenomenology (see, e.g. \cite{alvarez}) -- and
nonperturbative {\it quantum} frameworks are just on their way
\cite{bietenholz} -- hence, as in \cite{Castorina:2002vs}, we
consider the well behaved noncommutative {\it classical}
electrodynamics (NCED) proposed in \cite{jack2}. We shall take the
view that a nonzero $\theta$ -- the scale of noncommutativity --
breaks Lorentz invariance (c.f. e.g. \cite{Iorio:2001qy}), hence
$\theta$ ``measures'' the amount of Lorentz violation.

In \cite{Castorina:2002vs} large departures from the ordinary
spectrum of the synchrotron radiation were shown for the first
time. They are due to the acausal (Lorentz violating) behavior of
the electromagnetic fields related to the modified (shifted) poles
of the associated Green functions (see \cite{Castorina:2002vs},
\cite{ciz}, and Eqs. (\ref{a3fourier}) and (\ref{phifourier})
below). Those results pointed out the relevance of strong magnetic
fields of astrophysical origin to test Lorentz violation.

Here we want to pursue further that investigation. We do so
motivated by the recent result that in a Maxwell-Chern-Simons
electrodynamics there is Cherenkov radiation in vacuum
\cite{ralph}. We shall show that this also happens in NCED,
explicitly compute the spectrum of this radiation, and investigate
whether this novel feature might be useful to improve the bound on
$\theta$ by astrophysical limits on the Cherenkov radiation.

Noncommutativity for us is expressed in the canonical form
\cite{wess}, $x^\mu * x^\nu - x^\nu * x^\mu \equiv [ x^\mu , x^\nu
]_* = i \theta^{\mu \nu}$, where the $*$-product for $\phi(x)$ and
$\chi(x)$ is $(\phi
* \chi) (x) \equiv \exp\{ \frac{i}{2} \theta^{\mu \nu}
\partial^x_\mu \partial^y_\nu \} \phi (x) \chi (y)|_{y \to x}$,
$\theta^{\mu \nu}$ is $c$-number valued, the Greek indices run
from $0$ to $3$. The NCED action with a coupling to an external
current $J_\mu$ is
\begin{equation}  \label{ncym}
\hat{I} = - \frac{1}{4} \int d^4 x \hat{F}^{\mu \nu} \hat{F}_{\mu
\nu} + J_\mu \hat{A}^\mu \;,
\end{equation}
where $\hat{F}_{\mu \nu} = \partial_\mu \hat{A}_\nu - \partial_\nu
\hat{A}_\mu - i [ \hat{A}_\mu , \hat{A}_\nu ]_* $, $\hat{A}_\mu$
can be expressed in terms of a U(1) gauge field $A_\mu$ and of
$\theta^{\mu \nu}$ by means of the Seiberg-Witten (SW)
map\footnote{As $\theta^{\mu \nu} \to 0$, $\hat{A}_\mu (A,\theta)
\to A_\mu$ and $\hat{F}_{\mu \nu} \to F_{\mu \nu} =
\partial_\mu A_\nu -
\partial_\nu A_\mu$, hence this theory reduces to the ordinary
Maxwell theory, as requested.} $\hat{A}_\mu (A,\theta)$. The action
(\ref{ncym}) at $O(\theta)$ becomes
\begin{equation}  \label{othetamaxwell}
\hat{I} = - \frac{1}{4} \int d^4 x \; [F^{\mu \nu} F_{\mu \nu}
-\frac{1}{2} \theta^{\alpha \beta} F_{\alpha \beta} F^{\mu \nu}
F_{\mu \nu} + 2 \theta^{\alpha \beta} F_{\alpha \mu} F_{\beta \nu}
F^{\mu \nu}] + J_\mu \hat{A}^\mu \;,
\end{equation}
where we made use of the $O(\theta)$ SW map $\hat{A}_{\mu}(A,
\theta) = A_{\mu} - (1/2) ~ \theta^{\alpha
\beta}A_{\alpha}(\partial_{\beta}A_{\mu} + F_{\beta \mu})$, and of
the $*$-product. Our analysis is based on the theory
(\ref{othetamaxwell}).

We consider processes with emission of electromagnetic radiation
by external sources both in vacuum and in presence of an isotropic
medium with an electric permeability $\epsilon (\omega)$, that is
a function of the frequency of the radiation (the magnetic
permeability is taken to be 1). In this case the {\it linearized}
constitutive relations of \cite{jack2} among the fields in the
medium, descending from (\ref{othetamaxwell}), are only slightly
modified and are given by
\begin{equation}\label{linearized}
  D^i = \varepsilon^{i j} (\omega)  E^j \quad {\rm and} \quad H^i = (\mu^{-1})^{i j}
  B^j \;,
\end{equation}
where we choose $\theta^{0 i} = 0$, $\theta^{i j} = \epsilon^{i j
k} \theta^k$, $\epsilon^{i j k}$ is the completely antisymmetric
symbol, $\varepsilon^{i j} (\omega) = \epsilon(\omega) \xi^{i j}$,
\begin{equation}\label{epsmu}
  \xi^{i j} \equiv a \delta^{i j} +  \theta^i b^j + \theta^j b^i \; , \;
  (\mu^{-1})^{i j} \equiv a \delta^{i j} - (\theta^i b^j + \theta^j b^i) \;,
\end{equation}
$a = (1 - \vec{\theta} \cdot \vec{b})$, $\vec{b}$ is the
background magnetic field, and the Latin indices run from 1 to 3.
The Bianchi identities still hold unmodified, hence $F_{\mu \nu} =
\partial_{[\mu }A_{\nu]}$ or
\begin{equation}\label{potentials}
\vec{B} = \vec{\nabla} \times \vec{A} \quad {\rm and} \quad
\vec{E} = - \frac{1}{c} \frac{\partial}{\partial t} \vec{A} -
\vec{\nabla} \Phi \;.
\end{equation}
The dynamical Maxwell equations are $\partial_\mu \Pi^{\mu \nu} =
J^\nu + \theta^{\alpha \nu}  J^\sigma \partial_\alpha A_\sigma +
\theta^{\alpha \sigma}  \partial_\sigma (A_\alpha J^\nu)$, where
$\Pi^{\mu \nu} = \delta \hat{I} / \delta (\partial_\mu A_\nu)$,
and they lead to
\begin{eqnarray} \vec{\nabla} \cdot \vec{D} & = & 4 \pi
[\rho + \vec{\theta} \cdot
\left( \vec{\nabla} \times (\rho \vec{A}) \right)] \;, \label{maxdyn1} \\
\left( \vec{\nabla} \times \vec{H} - \frac{1}{c}
\frac{\partial}{\partial t} \vec{D} \right)^i & = & \frac{4
\pi}{c} \left[ J^i + \theta^j \left( \epsilon^{ijk} J^\sigma
\partial_k A_\sigma + \epsilon^{jlk} \partial_k (A^l J^i) \right)
\right] \label{maxdyn2} \;.
\end{eqnarray}

By using the potentials (\ref{potentials}), and the generalized
Lorentz gauge $\frac {\epsilon (\omega)}{c}
\frac{\partial}{\partial t} \Phi + \vec \nabla     \vec A = 0$,
the Euler-Lagrange equations (\ref{maxdyn1}) and (\ref{maxdyn2})
become
\begin{eqnarray}
a {\Box}_\epsilon \Phi + (\theta^i b^j + \theta^j b^i) \left[
\partial_i
\partial_j \Phi + \frac{1}{c} \frac{\partial}{\partial t}
(\partial_i A_j) \right] = -  \frac{4 \pi}{\epsilon (\omega)}
[\rho + \vec{\theta} \cdot
\left( \vec{\nabla} \times (\rho \vec{A}) \right)] \;, && \label{maxdyn4} \\
a {\Box}_\epsilon A^i + (\theta^i b^j + \theta^j b^i) \left[ -
\frac{\epsilon (\omega)}{c^2} \frac{\partial^2}{\partial t^2} A_j
+ \partial_j (\vec{\nabla} \vec{A}) \right] + \eps^{i k m}
(\theta^m b^j + \theta^j b^m)
\eps^{j l p} \partial_k \partial_l A_p & & \nonumber \\
= - \frac{4 \pi}{c} \left[ J^i + \theta^j \left( \epsilon^{ijk}
J^\sigma \partial_k A_\sigma + \epsilon^{jlk} \partial_k (A^l J^i)
\right) \right] \;, &&  \label{maxdyn3}
\end{eqnarray}
where ${\Box}_\epsilon \equiv -\epsilon (\omega) c^{-2}
\partial^2_t + \partial_{x_1}^2 +\partial_{x_2}^2
+\partial_{x_3}^2$, respectively.

In the same spirit of \cite{Castorina:2002vs}, we choose the
simplest possible settings to observe the effects of Lorentz
violation given by $\vec{b} = (0, 0, b)$ -- background magnetic
field speeding up the particle; $\vec{\theta} = (0, 0, \theta)$ --
i.e. $\theta^3$ is the only nonzero component of $\theta^{\mu
\nu}$. Although this choice for $\vec{\theta}$ surely is not the
most general, it has the double advantage for us of (a) capturing
the essential features of noncommutativity (see, for instance, the
acausal behavior of the electromagnetic field
\cite{Castorina:2002vs}) and (b) highly simplifying the Maxwell
equations (\ref{maxdyn4}) and (\ref{maxdyn3}). From these
settings: $\xi^{i j} = a \delta^{i j} + \delta^{i 3} \delta^{j 3}
\lambda$, where
\begin{equation}\label{lambdaref}
\lambda \equiv 2 \; \theta \; b \;,
\end{equation}
 $a = (1 - \theta b) = (1 - \lambda /
2)$. We are interested only in the largest departures from the
ordinary energy spectrum. These are $O(\theta)\cdot O(e^2)$ {\it
in the energy} -- where $e$ is the electric charge -- hence
contributions higher than $O(e)$ {\it in the fields} will be
neglected. Taking into account these approximations -- and the
further suppression at a large distance $R$ from the source, by a
factor $1/R$ -- Eqs. (\ref{maxdyn4}) and (\ref{maxdyn3}) can be
written as\footnote{For a more detailed analysis of the
approximations leading to Eqs. (\ref{approxa1})-(\ref{approxphi})
c.f. \cite{Castorina:2002vs}.}
\begin{eqnarray} {{\Box}_\epsilon} {A}_1 + \lambda
\partial_2 (\partial_1 {A}_2 -
\partial_2 {A}_1) & = & - \frac{4 \pi}{c} \tilde{J_1} \;, \label{approxa1} \\
{{\Box}_\epsilon} {A}_2 + \lambda \partial_1 (\partial_2 {A}_1 -
\partial_1
{A}_2) & = & - \frac{4 \pi}{c} \tilde{J_2} \;, \label{approxa2} \\
{{\Box}_\epsilon} {A}_3 - \frac{\lambda\epsilon (\omega)}{c^2}
\partial^2_t {A}_3 - \frac{\epsilon (\omega)}{c} \lambda \partial_3
\partial_t
\Phi & = & - \frac{4 \pi}{c} \tilde{J_3} \;, \label{approxa3} \\
{{\Box}_\epsilon} {\Phi} + \lambda ( \partial^2_3 \Phi +
\frac{1}{c}
\partial_3 \partial_t A_3) & = & - \frac{ 4 \pi \tilde{\rho}}{\epsilon (\omega)} \label{approxphi} \;,
\end{eqnarray}
where $\tilde{J}_i \equiv J_i / a$, $i = 1,2$, and $\tilde{\rho}
\equiv \rho / a$. In the following Section we shall present the
computation of the vacuum Cherenkov radiation in these settings,
while the last Section is devoted to the discussion of some
astrophysical implications and to our conclusions.

\section{Vacuum Cherenkov radiation in NCED}

In our settings $\vec{\theta}$ only has a $z$-component, hence the
largest modifications of the Cherenkov spectrum are expected when
the charged particle moves along the same axis:
$\vec{\tilde{J}}=(0,0,\tilde{J}_3)$. As Eqs. (\ref{approxa1}) and
(\ref{approxa2}) represent the propagation of plane waves in NCED
discussed in \cite{jack2}, the equations of interested for the
present analysis are Eqs. (\ref{approxa3}) and (\ref{approxphi})
that, in momentum space, have the following solutions
\begin{eqnarray}
A_3 (\vec{k}, \omega) & = & - \frac{4 \pi}{c} \frac {\tilde{J}_3
(\vec{k}, \omega)} {(1+\lambda) \epsilon (\omega) \frac{\omega^2}{
c^2} - \vec{k}^2 } - \frac{4 \lambda \pi}{c} \frac{\tilde \rho k_3
\omega} { [(1+\lambda) \epsilon (\omega) \frac{\omega^2}{c^2} -
\vec{k}^2 ] (\epsilon (\omega)\frac{\omega^2}{c^2}  - \vec{k}^2
-\lambda k_3^2 )}
\;, \label{a3fourier} \\
\Phi (\vec{k}, \omega) & = & - \frac{ 4 \pi \tilde{\rho} (\vec{k},
\omega) / \epsilon (\omega)} {\epsilon (\omega)
\frac{\omega^2}{c^2} - \vec{k}^2 - \lambda k_3^2} + \frac{4
\lambda \pi}{c^2} \frac{ k_3 \omega   \tilde{J}_3   } { [
(1+\lambda) \epsilon (\omega) \frac{\omega^2}{c^2}  - \vec{k}^2 ]
(\epsilon (\omega) \frac{\omega^2}{c^2} - \vec{k}^2 -\lambda k_3^2
)} \label{phifourier} \;,
\end{eqnarray}
where $\tilde{\rho} (\vec{k},\omega)= e / (2 \pi
a)~\delta(\omega-k_3 v)$, $ \tilde{J}_3 (\vec{k}, \omega)=(0,0,v
\tilde{\rho})$ and $v$ is the speed of the particle.

If one defines \be \vec E(\omega) =\frac{1}{(2\pi)^{3/2}}\int d^3
k \vec E(\vec k, \omega) e^{i w k_2} \label{eomega} \ee where $w$
is the perpendicular distance from the path of the particle moving
along the $z$-axis and the observation point of  $\vec E(\omega)$
has coordinates $(0,w,0)$, the energy per unit distance lost in
collisions with the medium with impact parameter $w \geq d$ --
where $d$ is a large distance from the path -- is seen to be given
by \cite{jackson} \be \frac{d {\cal E}}{d z}= - {\rm Re}\left ( i
\int_0^\infty d\omega ~ \omega \epsilon (\omega) {\cal E} (\omega,
\lambda, d) \right) \;,\label{enloss} \ee where
\begin{equation}\label{F(E)} {\cal E} (\omega, \lambda , d) \equiv
\int^{\infty}_d dw ~w~ | \vec E(\omega)|^2 \;,
\end{equation}
and we wrote explicitly the $\lambda$-dependance. For us the
relevant components of the electric field are \be E_2(\omega)=
\left ( \frac{2}{\pi}\right )^{1/2} \frac{e}{v (1+\epsilon
(\omega) \beta^2)} \left \lbrack \frac {\sigma_+
K_1(w\sigma_+)}{\epsilon (\omega)} + \beta^2 \sigma_-
K_1(w\sigma_-) \right  \rbrack \;, \label{e2} \ee and \be
E_3(\omega)= \left ( \frac{2}{\pi}\right )^{1/2} \frac{- i e
\omega (1- \epsilon (\omega)\beta^2) }{v^2 (1+ \epsilon
(\omega)\beta^2) } \left \lbrack \frac { K_0(w\sigma_+)}{\epsilon
(\omega)} + \beta^2 K_0 (w\sigma_-) \right \rbrack \;, \label{e3}
\ee where $\beta = v / c$, \be \sigma_+^2=\frac{\omega^2}{v^2}
\left \lbrack 1+\lambda -\epsilon (\omega) \beta^2 \right \rbrack
\quad \;, \quad \sigma_-^2=\frac{\omega^2}{v^2} \left \lbrack 1 -
( 1+ \lambda) \epsilon (\omega) \beta^2 \right \rbrack \;,
\label{sigmas} \ee and $K_n$ are the modified Bessel functions.
Inserting the expressions for the fields (\ref{e2}) and (\ref{e3})
into Eq. (\ref{F(E)}), we see that for large $d$, one can use the
asymptotic behavior of $K_n$ to write \bea {\cal E}(\omega,
\lambda, d) & = & \frac{e^2}{v^2} \left \lbrace \frac{|A |^2+ |C
|^2 \omega^2/v^2 }{ 2 ~ |\sigma_+| ~ {\rm Re}(\sigma_+)} \; \;
\exp (- 2 d ~ {\rm Re}(\sigma_+)) + \frac{|B|^2+ |D|^2
\omega^2/v^2 }{2 ~ |\sigma_-| ~ {\rm Re}(\sigma_-)} \; \; \exp (-
2 d ~ {\rm Re}(\sigma_-) )\right.
\nonumber\\
&+& \left. 2 {\rm Re} \left ( \frac  { A  B^* +  C D^*
\omega^2/v^2  } { \sigma_+^2 - {\sigma^*_-}^2 } ~ \frac{ \sigma_+
- \sigma_-^* }{\sqrt{ \sigma_+ \sigma^*_-}} \; \exp(- d ~ (
\sigma_+ + \sigma^*_- )) \right ) \right\rbrace \;, \label{abcd}
\eea where \be
 A = \frac{ \sigma_+}{\epsilon (\omega) ( 1+\epsilon (\omega) \beta^2 )}
\;, \; B = \frac{ \beta^2 \sigma_-} {1+\epsilon (\omega) \beta^2}
\; , \; C =  \frac{1-\epsilon (\omega) \beta^2}{\epsilon (\omega)
(1+\epsilon (\omega) \beta^2 )}\; , \; D = \frac{ \beta^2 (
1-\epsilon (\omega) \beta^2 )}{1+\epsilon (\omega) \beta^2} \;.
\ee

For $\lambda = 0$ we have $\sigma_+ = \sigma_- \equiv \sigma$,
thus the correct Maxwell limit is recovered because, for large
$d$, ${\cal E} (\omega, 0, d) \sim \exp(- 2 d ~ {\rm Re}
(\sigma))$, hence the Cherenkov radiation is observed only if
${\rm Re} (\sigma) = 0$, which never happens in vacuum - {\it in a
medium} this is the usual condition that gives $v > c$ (c.f.
either of Eqs.(\ref{sigmas}) with $\lambda =0$).

In NCED, instead, Cherenkov radiation is also allowed {\it in
vacuum}. Indeed, from Eqs. (\ref{sigmas}) we can have either ${\rm
Re} (\sigma_+) = 0$ or ${\rm Re} (\sigma_-) = 0$, for
\begin{equation}\label{limit1}
\lambda < \beta^2 - 1 \quad {\rm and} \quad \lambda < 0 \;,
\end{equation}
or \be \lambda
> \frac{1-\beta^2}{\beta^2} \quad {\rm and} \quad \lambda > 0 \;, \label{limit2} \ee
respectively, where $\lambda = 2 \theta b$, hence the dumping at
large $d$ in Eq.(\ref{abcd}) could, in principle,
disappear\footnote{The real part of the third exponential $\exp(-
d ~ ( \sigma_+ + \sigma^*_- ))$ in Eq. (\ref{abcd}) can never be
$1~$ for $\lambda \neq 0$.}. It makes sense to investigate only
the behavior in the ultrarelativistic regime: $\beta^2 = 1 -
\eta$, where $\eta$ is a small positive number, by keeping only linear
terms in $\eta$. To this order  both conditions (\ref{limit1})
and (\ref{limit2}) amount to $|\lambda| > \eta$.
We can then pick one of them, say (\ref{limit2}), to
write the energy loss in a NCED vacuum given in Eq.(\ref{enloss})
as\footnote{As customary, we used $\epsilon = 1 + i {\rm Im}~
\epsilon$ and only at the end set ${\rm Im}~ \epsilon = 0$.}
\be \frac{d {\cal E}}{d z}= \int_0^\infty d\omega ~ \omega ~
\frac{(v^2/\omega^2) |B |^2+ |D |^2}{(1+\lambda) \beta^2} ~ e^{- 2
d ~ {\rm Re}(\sigma_-)} = \int_0^\infty d\omega \frac{ \omega e^2
\beta^2  }{c^2}~ \frac{\lambda-1+\beta^2}{(1 +\beta^2)^2}
\label{vaccrkv1} \;,\ee
where we explicitly wrote $1 = \exp (- 2 d ~ {\rm Re}(\sigma_-))$
to make clear the connection with Maxwell theory.
By neglecting $O(\eta^2)$ and $O(\eta\lambda)$ contributions,
Eq. (\ref{vaccrkv1}) reads
\begin{equation}\label{vaccrkv2}
    \frac{d {\cal E}}{d z} \sim \frac{1}{4} \int_0^\infty d\omega \frac{ \omega
    e^2}{c^2}~ (\lambda - \eta) \;.
\end{equation}
The energy loss corresponding to the other condition, i.e. to  Eq. (\ref{limit1}),
in the same approximation, is given again
 by Eq. (\ref{vaccrkv2}),  but with $\lambda$ replaced by $|\lambda|$.
Finally the same conditions, Eqs. (\ref{limit1}) and
(\ref{limit2}), can be recovered by the expression of the emission
angle of the radiation (see \cite{jackson}).

\section{Phenomenological Implications and Conclusions}

Let us summarize here our understanding of the state of the art:

In \cite{Castorina:2002vs} the departures from the ordinary
spectrum of the synchrotron radiation due to Lorentz violating
(acausal) behaviors of the electromagnetic fields gave a large
correction factor $X$ to the energy emitted by the
source\footnote{Similar large corrections appear to be reobtained
in \cite{Montemayor:2004mh} for the synchrotron in Myers-Pospelov
effective electrodynamics.}
\begin{equation}
X <  (\frac{\omega_0}{\omega})^{2/3} n \times 10^{-21} \times
\left( \frac{{\cal E} ({\rm MeV})}{{\rm MeV }}\right)^{4} \;,
\label{sinc}
\end{equation}
where this formula holds in the ultra-relativistic approximation
for $\omega_0 << \omega << \omega_c$, $\omega_0$ is the cyclotron
frequency ($\omega_c = 3 \omega_0 \gamma^3$), $\gamma = (1 -
\beta^2)^{-1/2}$, $n$ is the value of the background magnetic
field $b$ expressed in Tesla, ${\cal E}$ is the energy of the
source and we used the bound\cite{carroll} $\theta < 10^{-2} ({\rm
TeV})^{-2}$. From (\ref{sinc}) it was clear that, in spite of the
$\gamma^4 $ correction, only very strong magnetic fields $b$ could
improve the current bound on $\theta$. Hence strong magnetic
fields of astrophysical origin were seen to be the proper
candidates to test Lorentz violation in this framework. The
generality of this result resided in the poles-shift mechanism
that one has to expect in a Lorentz violating electrodynamics.
This point was also addressed in \cite{ciz}.

In \cite{jack2} it was shown that plane waves in NCED have a
deformed dispersion relation $\omega/c = k (1 - \vec{\theta}_T
\cdot \vec{b}_T) $. This modifies the kinematical thresholds of
processes involving radiation, such as $\gamma \gamma \to e^+
e^-$, and otherwise forbidden decays, such as $\gamma \to e^+
e^-$, are permitted. The astrophysical observation of ultra high
energy gamma rays put some tight bounds on the modification of
these thresholds in any theory involving violation of Lorentz
invariance \cite{jackli}, \cite{coleman} (for a review see e.g.
\cite{varie}). In NCED these kinematical arguments do not improve
the present bound on $\theta$ as the galactic and extragalactic
magnetic field is too weak \cite{noiepl}. As already noted in
various occasions, however, a sound quantum version of NCED is not
available. Thus, the considerations about such radiation processes
only refer to the modified mass-shell (or light cone) of the
theory.

Let us now consider the possibility of a phenomenological bound on
$\theta$ coming from the Cherenkov radiation in the NCED vacuum
just obtained in the previous Section. The kinematical analysis of
vacuum Cherenkov radiation, performed in other Lorentz violating
theories of electrodynamics, also introduces strong limits on
Planck scale \cite{steck}. The 50 TeV gamma rays experimentally
seen from the Crab nebula should come from highly energetic
electrons which are explained by inverse Compton scattering. But
this rules out the vacuum Cherenkov radiation because the
Cherenkov rate is orders of magnitude higher than the inverse
Compton scattering rate \cite{coleman}, \cite{steck}.

For us this implies that the conditions in Eqs. (\ref{limit1}) and
(\ref{limit2}) are not fulfilled, which in turn produces a limit
on $\theta$. However, even  by considering electrons of very high
energy (such as 1.5 PeV \cite{steck}) with the present bound on
$\theta$ one obtains the condition $n ~10^{-23} < 10^{-19}$, which
is largely satisfied. Thus tight bounds could only be obtained by
considering very strong magnetic fields, such as those observed in
compact stars. Since the production of high energy gamma rays from
e.g. pulsars, is a complex multi-step process involving shock
waves, primary and secondary emissions, synchrotron radiation,
inverse Compton scattering, and more \cite{ho}, one should
evaluate how and how much each of those steps is modified.
Numerical evaluations might be a conceivable approach to this
involved analysis.

\vspace{.5cm}

\noindent {\bf \large Acknowledgments}

\noindent We thank Roman Jackiw for useful and enjoyable
discussions. A.I. thanks A.P. Balachandran for a discussion on
unitarity of noncommutative field theories. This work is supported
in part by funds provided by the U.S. Department of Energy
(D.O.E.) under cooperative research agreement DF-FC02-94ER40818.

\end{document}